# Vertically aligned graphene based non-cryogenic bolometer

*Kiran Shankar Hazra*[1,3*], *N. Sion*[2], *Anil Yadav*[1], *James McLauhglin*[2] and *Devi Shanker Misra*[1]

[1]Department of Physic, Indian Institute of Technology Bombay, Mumbai-400 076 (India)

[2]Nanotechnology and Integrated BioEngineering Centre, University of Ulster, Jordanstown, Newtownabbey, Northern Ireland, BT37 0QB (UK)

[3]Physics & Astronomy Department, University of California, Los Angeles, CA-90025 (USA)

## Abstract

We report the photoresponse of vertically aligned graphene upon IR irradiation at room temperature. Four probe measurements have shown electrical switching in I-V characteristics during pulsed IR irradiation. The photoresponse reported here for vertically aligned graphene (VAG) is much higher than carbon nanotube (CNT) samples. Our investigation has shown that such photoresponse arise solely due to bolometric effect, where the conductivity changes with temperature. The magnitude of the resistance of VAGs increases by ~ 2 fold for 6 $^0$C increase in temperature. Also the Thermal Coefficient of Resistance (TCR) in this region is ~11%/K, which is the highest TCR value reported so far for any carbon nanomaterials.

*Corresponding author: Dr. K. S Hazra

kiran@physics.ucla.edu

Among the carbon materials, carbon nanotubes (CNTs) have already been proven to be very useful material for infrared photoresponse. Several reports have confirmed that both the SWNTs and MWNTs have obvious response to IR irradiation. For instance, Itkis *et al.* have shown sensing of IR by a suspended CNT network in cryogenic temperature in vacuum [1]. IR photo response has also been reported for individual isolated CNTs or CNT-polymer composites even at room temperature and normal pressure [2-4]. However the origin of photo-response in CNTs has always been debated as to whether it is due to (1) photo induced exitons, (2) interband transition or (3) bolometric effect. M. Freitag et.al.[5] had shown that IR radiation can generate excitons or bound electron hole pairs in CNTs which can be separated by applying electric field and can contribute to photo-response. M.E.Portnoi et al.[6] theoretically predicted that IR can be absorbed or produced due to interband transition, generating free electron hole pairs in CNTs. Results also have shown that IR raditaion can be absorbed by the direct band gap of semiconducting CNTs[7]. However the majority of the reports suggest that the photo-response of CNTs arises due to bolometric effect, where the conductivity of CNTs decreases with temperature due to electron-phonon interaction [1, 8, 9]. For semiconducting CNTs, increase in conductivity due to thermally generated carrier concentration is also possible in bolometric effect [2, 4]. The bolometric response is not very strong for CNT samples in normal condition, thus efforts have also been made to improve the performance of bolometer by using cryogenic stage, polymer composite with CNTs or using vacuum, which reduces the thermal cross-linking of samples [1, 3].

In this paper we have thoroughly investigated the IR photoresponse of Vertically Aligned Graphene (VAG) at normal room temperature and have reported enhanced IR photo-response as compared the reports on CNTs. Graphene has already been proven to be very promising nanomaterial for applications in nanoelectronics, nanofillars etc. like its ancestor, carbon nanotube [10-17], however its photoresponse character is yet not well established. Here we have reported a swift electrical switching behavior due to photoresponse of the graphene samples in pulsed IR irradiation. Our investigations indicate that the photo-response in our graphene samples arise solely due to bolometric effect and not due to excitons or interband transition photoconductivity. The advantage of graphene as compared to CNTs is their higher surface area and unique band structure. Using vertically aligned grapheme (VAG) sample we have shown considerably higher response to IR irradiation as compared to CNTs.

In figure 1(a) shows the scanning electron microscopy (SEM) images of as the grown graphene samples. It is evident that the samples are almost vertically aligned with the graphene platelets having curved structure intercalated randomly to each other and form a mesh like network. The sizes of the platelets are found varying from ~100 nm to ~2 µm. Platelet tip consist of 5-10 layers of graphene and the base of the platelets are quite thicker perhaps consisting of 10-20 layers. The Raman spectra shown in figure 1(b) confirm the graphene structure with a strong 2D peak at 2700 cm$^{-1}$. The intensity ratio of G-band and D-band is ~4, implies a well-defined crystalline structure of the VAGs. Since the graphene platelets are vertically aligned in nature the the Raman signal mostly originates from the edges of the graphene, which could be the reason for such intense 2D peak in Raman spectrum.

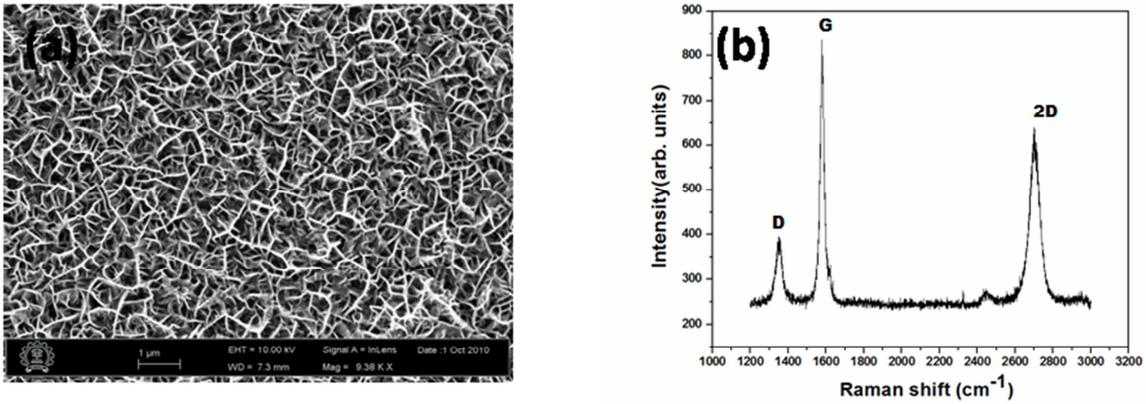

Figure 1: (a) SEM images of the graphene sample, showing intercalated graphene flakes, entangled in vertical direction to the sample surface. (b) Raman spectra of the VAGs shows sharp G and 2D peaks, including a low intensity D peak.

The bolometric measurements were done on the VAG graphene film at room temperature (293 K). The schematic diagram of the experimental set up for the bolometric measurements is shown in figure 2. Silver paste was used to improve the contact of the probes with the VAGs. We have used a 200 W IR lamp to irradiate the VAGs and it was placed at a distance 30 cm on top of the sample. The incident IR radiation power on the sample was ~6 mW/mm$^2$ measured by calorimetric technique. We have used a Cr-Al thermocouple to measure the real time temperature of the VAG sample. The thermocouple was directly in contact with the VAGs instead of placing it in sample base, which is very novel in our process. Here we have done simultaneous measurements of the resistance change and the temperature response of the VAGs.

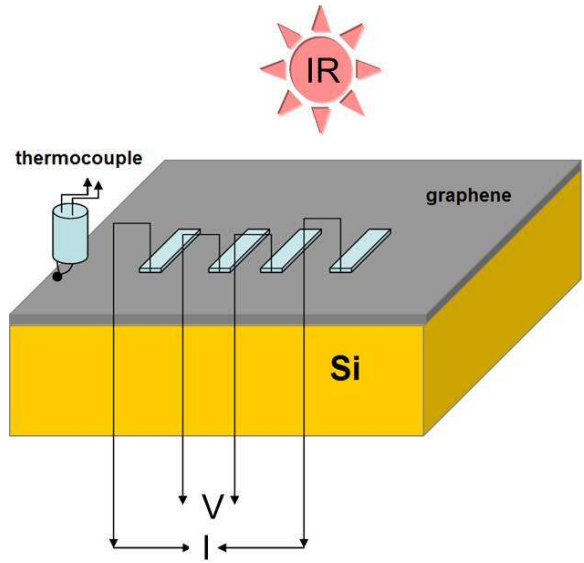

Figure 2. Schematic diagram of the experimental set up.

In figure 3a we have shown the I-V characteristics of the sample in normal conditions (at 293 K normal day light) and upon IR irradiation. In normal condition the I-V shows perfect linear characteristics, which confirms existence of perfect ohmic contact on the graphene sample and having no change in contact resistance throughout the experiments. The resistance recorded at normal condition was ~ 1.8 k-ohm. To examine the photo-response of the VAG sample we have measured the voltage response upon periodic irradiation of IR on the sample at a constant current (~ 2 mA) in the circuit. In figure 3 we have shown the switching characteristics of photoresponse due to periodic IR irradiation having 60 sec pulse with 90 sec interval. Such long duration pulses were kept to record the saturation and the baseline behavior of the response. When the IR was switched on the resistance increases rapidly and goes to saturation and when the IR is switched off the resistance again goes to baseline with a sharp drop. The plot shows that the baseline and the saturation line are almost parallel, indicating the changes are solely due to photo excitation and not due to any change in contact resistance. The response time to reach the resistance from the baseline to saturation is ~ 5 sec. Such high response time and increase in resistance upon IR irradiation can occur only due to bolometric photo-response [8, 9, 18, 19]. Whereas for the photoresponses due to excitons or due to photoconductivity for interband transition, the response time would be much lower (ranges within $10^{-9}$ sec to $10^{-15}$ sec) and the resistance decreases upon IR irradiation. The normalized resistance $R/R_1$ (R is the actual resistance and $R_1$ is the maximum resistance during saturation) raised ~ 2 fold in the bolometric measurement at room temperature, which is quite remarkable. The sensitivity of the bolometer is defined as, $s = \frac{R_{sat} - R_{base}}{R_{base}} \times 100\%$, where $R_{sat}$ is the saturation resistance and $R_{base}$ is the base line resistance. In most of the previous reports on photoresponse of CNTs, the sensitivity has been found within the range 0.01% to 30%, even some of them were measured in cryogenic temperature [1, 3, 4]; whereas, in our case we report ~ 100% increase in resistance due to IR irradiation at room temperature, despite of using similar radiation power as it was used in the previous reports on CNT bolometric response. Such high range in resistance variation due to bolometric response can facilitate detection of very small difference in temperature, which could be very useful for new generation bolometric device applications.

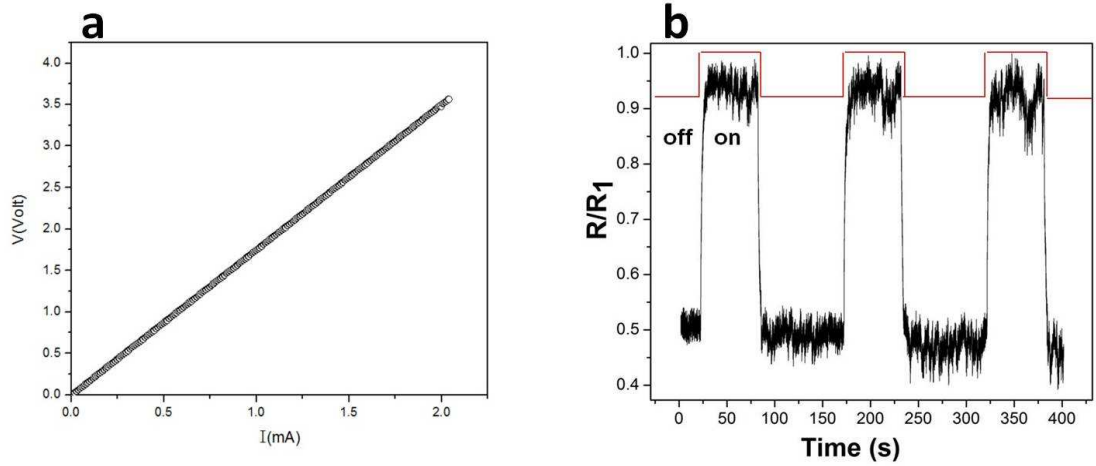

Figure 3: (a) I-V characteristics of VAGs at normal condition. (b) Electrical switching characteristics for pulsed IR irradiation.

To investigate the behavior of the bolometric photoresponse of the VAG sample thoroughly we have recorded the resistance change for different pulse duration (5 s, 10 s, 20 s, 30 s, 40 s and 50 s) of IR irradiation. Figure 4 shows the changes in normalized resistance for different pulse durations and the corresponding temperature of the VAGs, measured by the thermocouple, which was directly in contact with the VAG flakes. For the 5 s pulse, the normalized resistance just reaches the saturation region, whereas for higher time duration pulses the saturation regions were well established. It was found from each plot that the normalized resistance varies with time (t) exponentially during heating and cooling as $\frac{R}{R_1} = u - v \exp(-w \times t)$ and $\frac{R}{R_1} = A + B \exp(-C \times t)$, respectively with u, v, w and A, B, C are positive constants. Also with higher duration IR pulses (30 s onwards) it can be clearly noticed that resistance periodically fluctuates at saturation region which may be due to the periodic fluctuation of temperature at steady state.

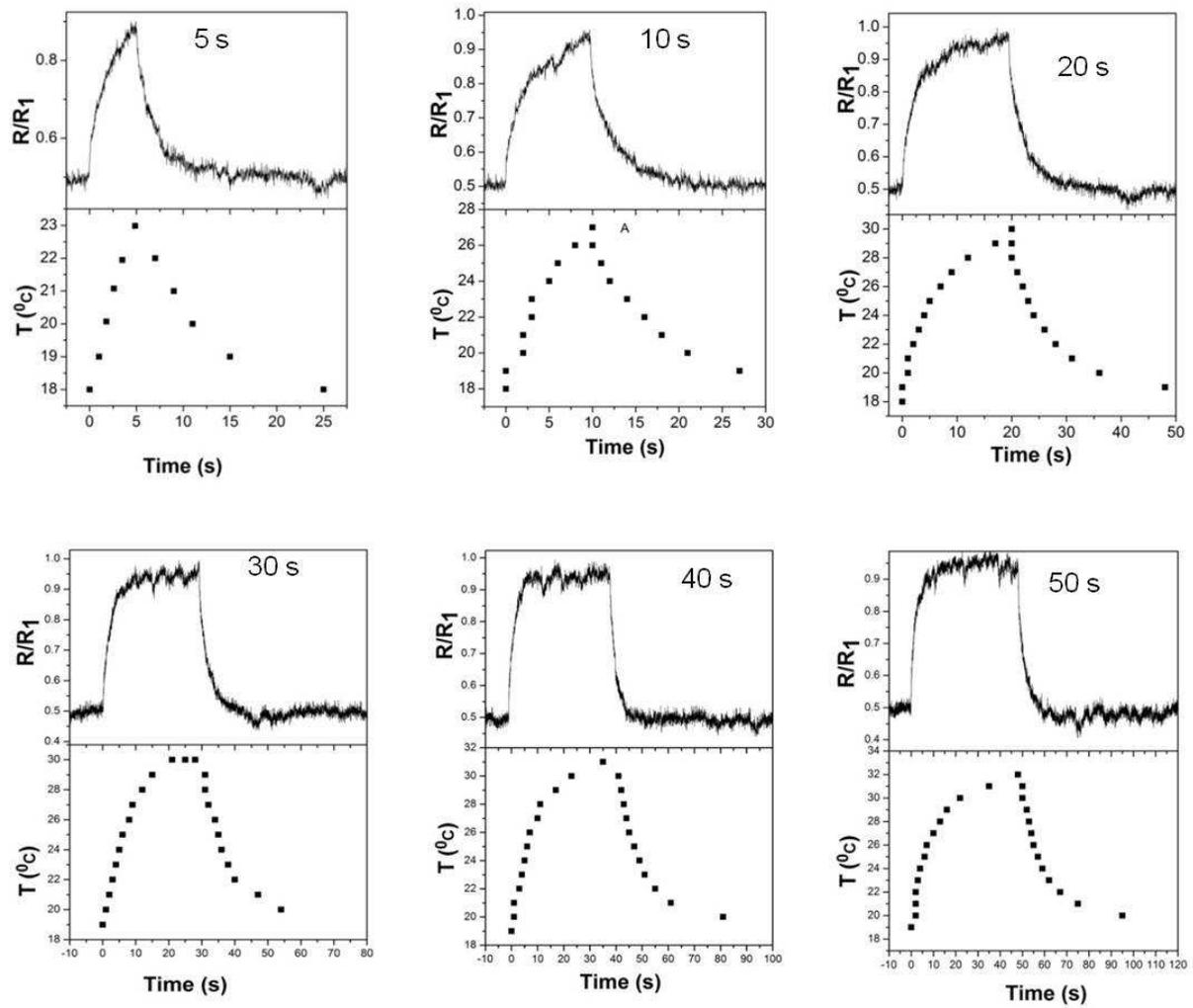

Figure 4: Change in normalized resistance and temperature of the VAGs with time for different duration (5 s, 10 s, 20 s, 30 s, 40 s and 50 s) IR pulses.

It is very clear from the figure that the resistance varies in similar fashion as the temperature reading of the thermocouple, confirming our argument to indentify the present photoresponse as bolometric. According to the thermocouple reading the steady state temperature is ~ $(31 \pm 1)$ $^0$C. The major difference between the temperature curve and the resistance curve is that the resistance curve approach towards the saturation much faster than the temperature curve, indicating higher response of the VAG bolometer than the thermocouple. To get a clear view of the temperature dependence of VAGs resistance we have plotted the hysteresis loop (figure 5) of normalized resistance w.r.t the thermocouple reading for 5 s IR pulse. Here the 5 s pulse was chosen to avoid the steady state saturation region and to get the response only during cooling and heating. At room temperature both the heating and cooling curve shows nonlinear response of resistance to temperature, although at low temperature and for short range temperature change the response may be considered as linear. During heating the normalized resistance rises rapidly with temperature at low temperature region and as the temperature increases the rate of increase becomes slower and approaches towards steady state; it follows the exponential function $\frac{R}{R_1} = \alpha - \beta \exp(-\gamma \times T)$, where $\alpha = 0.87$, $\beta = 322887$, $\gamma = 0.63$ and T is the temperature in $^0$C. The scenario reverses during cooling and it obeys the function $\frac{R}{R_1} = x + y \exp(z \times T)$, where x = 0.50, y = 8.9 x $10^{-12}$ and z = 1.06. The change in resistance with temperature, $\frac{dR}{dT}$ is positive for heating curve and it shows negative value for cooling curve, which is expected for bolometric photo-response and it indicates metallic behavior of the VAGs. One of the important measuring parameter for bolometric response is the Thermal Coefficient of Resistance (TCR) and it is defined as, $= \frac{1}{R}\frac{\Delta R}{\Delta T}$, which is the normalized change in resistance with temperature. The TCR of the VAG bolometer can be calculated from the heating curve of hysteresis loop; while temperature rises from 18-23 $^0$C, the TCR is ~11%/K, which is quite high TCR value as compared to previous reports of TCR on CNTs showing range within 0.01-3 %/K [1, 3, 4]. TCR can show higher values for lower temperature range as the temperature goes away from the steady state temperature and the slope of the heating curve increases. The reason for getting such high TCR in our case is probably the high surface area of the graphene as compared to CNTs. The flux of IR radiation will be much higher for graphene samples, especially for VAGs (due to their alignment), as compared to the CNTs. It could also be anticipated from previous reports[1, 3] that the TCR could be improved further at cryogenic temperature or arranging thermal isolation through vacuum for the VAG sample, which could pave a path towards next generation bolometric sensors.

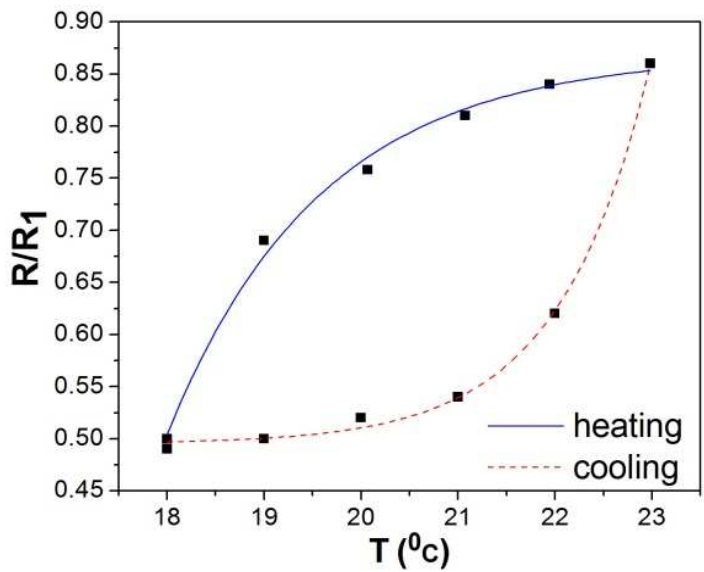

Figure 5: Bolometric hysteresis loop for normalized resistance of VAGs.

In conclusion the VAG sample shows significant photoresponse as compared to CNTs. The electrical conductivity switching of I-V characteristics has been noticed upon pulsed irradiation of IR. Two fold increases in resistance was noticed from baseline to saturation region, which is due to bolometric effect. The resistance of the sample varies in a similar fashion as the temperature of the sample. The hysteresis of the VAG's resistance shows TCR is ~11%/K for the temperature rising from 18 $^0$C to 23 $^0$C, which is the highest value, recorded for carbon nanomaterials. Also large surface area, catalyst free synthesis technique and strong adhesion to the substrate gives extra advantage to VAGs for bolometric applications.

**Experimental:**

**Graphene preparation:**

The synthesis of VAG samples was carried out in a SEKI Microwave Plasma Enhanced Chemical Vapour Deposition (MPECVD) deposition system. The system used a 1.5 kW, 2.45 GHz microwave power source. Heavily doped Si substrates were initially heated in-vacuum up to 900°C for the cleaning purpose and then subjected to nitrogen plasma at 650 Watts power and 40 Torr. The growth of VAGs was initiated by introducing $CH_4$ into the chamber with a $CH_4/N_2$ gas flow in the ratio of 1:4. During the growth the microwave power was raised to 800 W and maintained during the growth time of 60 sec. The samples were then allowed to cool down under a constant $N_2$ flow.

**Instrumental:**

Four probe measurements were carried out on the VAG sample upon IR irradiation using Keithley 220 current source and Keithley 2182 nano-voltmeter. A confocal micro-Raman spectrometer (Jobin Yvon, model: HR800) was used to characterized CNTs and graphene at room temperature. It uses 20 mW, 514.5 nm $Ar^+$ laser with typical resolution of 0.5 $cm^{-1}$. SEM and HRTEM was carried out in Jeol-6390 and JSM-2010F set ups.